# The atom and atom scattering model to control ultracold collisions

Hasi Ray[1,2]

[1]Department of Science, National Institute of Technical Teachers' Training and Research Kolkata Block FC, Sector 3, Salt Lake City, Kolkata 700106, India
[2]Department of Physics, New Alipore College, Kolkata 700053
Email: hasi_ray@yahoo.com

**Abstract**: A modified static exchange model is introduced, it could be useful to control the elastic s-wave scattering length in studying cold-atomic interaction. The theory includes the long-range van der Waals interaction in addition to short-range exchange force. The model is applied on Ps-H system to study the Feshbach resonances e.g. the s-wave elastic phase shifts and the corresponding cross sections at ultracold energies. These are compared with the data obtained using the static exchange model. A very interesting features are observed when we vary the interatomic separation i.e. the strength of dipole-dipole interaction.

**Introduction:**
Experimental realization of Bose-Einstein condensation (BEC) in dilute gases of alkali-metal atoms [1] and hydrogen atoms [2] has generated interest in ultracold physics [3]. The controlling of the s-wave scattering lengths in ultracold collisions is an important subject of study to tune the Feshbach resonances. In ultracold systems close to BEC, the kinetic energies of the atoms are negligibly small and the interaction time is much longer than normal atomic interactions. If the density is $\sim 10^{10}$ atoms/cm$^3$; the interatomic separation is $\sim 10^{-4}$ to $10^{-3}$ cm, it is $10^4$ to $10^5$ times larger than atomic dimension $\sim 10^{-8}$ cm [4]. Two of the slowly moving atoms can come close to each other when all others are far apart. In this approximation, the atomic collision physics can provide reliable information about the cold-atomic system [5]. The electron-electron exchange correlation and the dipole-dipole long range interaction forces start dominating as the system moves towards colder energy region. The static exchange model [6,7] includes the electron-electron exchange correlation force exactly to solve Schrodinger equation, but no long-range interaction and no channel coupling. We introduce a modified static exchange model (MSEM) in which attempts are made to include both these effects to study cold-atomic interaction and Feshbach resonances. In optical lattices where the ultracold atoms are confined in the centre of the trap, we can suppose that there is no effect of magnetic field used to trap the atoms. So the present model (MSEM) can be useful to control the s-wave scattering lengths. The same type of approximation was earlier used by Barker and Bransden to study the quenching of orthopositronium by helium [8].

A large group of scientists are involved in studying the magnetically tunable scattering resonances [9] in ultracold atomic collisions and a few are involved in studies to tune the system using a dc-electric field [10]; generally a magnetic trap is used to confine the atoms and hyperfine transitions of atomic energy levels for laser cooling [11,12]. The cause of BEC at ultracold temperatures is supposed to be due to the attractive long-range interaction between the atoms. The well-known long-range interaction between two atoms are the van der Waals interaction which arises due to dipole-dipole interaction. It is defined as $V_{van}(R) = -\frac{C_W}{R^6}$, where $C_W$ is the v*an der Waals coefficients* and $R$ is the interatomic separation. Barker and Bransden in ref. [8] calculated the van der Waals interaction between the positronium and helium atoms, it has the form :

$$V_{van}(R) = 0, \quad \text{if} \quad R \angle R_0 \quad \text{and}$$

$$V_{van}(R) = -\frac{C_W}{R^6}, \quad \text{if} \quad R \geq R_0 \quad \text{when} \quad R_0 \to 0.$$

The similar concept is discussed in Lennard-Jones 6-12 potential [13]. It is defined as

$$V_{LJ}(R) = -4\varepsilon\left\{\left(\frac{\sigma}{R}\right)^6 - \left(\frac{\sigma}{R}\right)^{12}\right\}$$

when $\varepsilon$ and $\sigma$ are *the Lennard-Jones parameters.*

In the above expression, the first term corresponds to the van der Waals interaction and the second term represents the short range electron-electron exchange correlation force and it is repulsive. In principle, the short range electron-electron exchange force is repulsive if the electron spins form a triplet state, it is attractive if the spins form a singlet state.

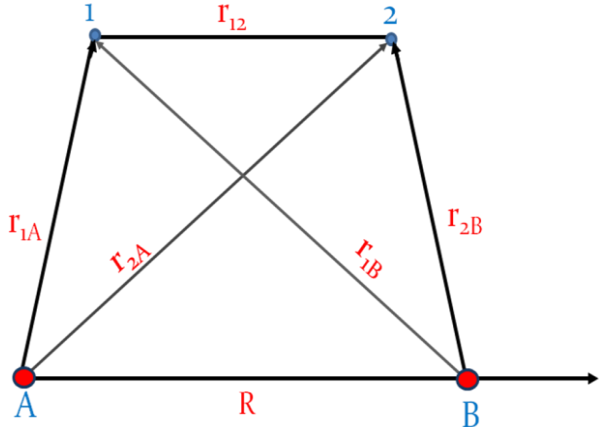

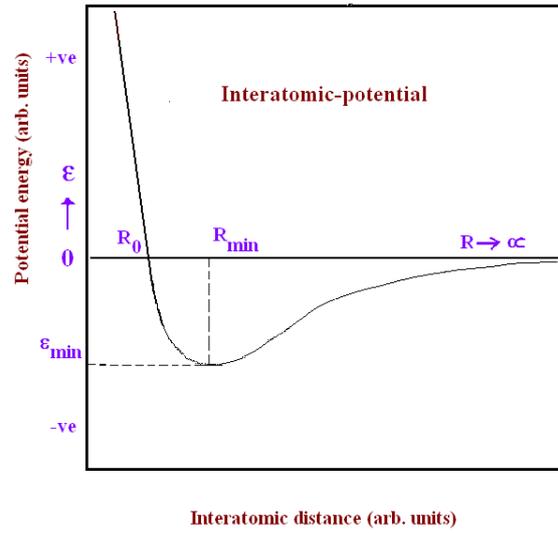

Figure 1                                                   Figure 2

**Theory:**
In the static approximation, the Schrodinger equation

$$H\psi = E\psi$$

is solved using the eigen state expansion methodology

$$\psi = \sum_i a_i \varphi_i$$

with only the direct elastic channel. The elastic channel is defined so that $\left|\vec{k}_i\right| = \left|\vec{k}_f\right|$; here $\vec{k}_i$ and $\vec{k}_f$ are the initial and final momenta of the projectile.

In the static exchange model (SEM), we include the effect of exchange or antisymmetry of the system electrons. The initial and final channel wave functions are defined as

$$\psi_i = e^{i\vec{k}_i \cdot \vec{R}} \phi_{1s}(r_{1A})\phi_{1s}(r_{2B})$$

$$\psi_f = (1 \pm P_{12})e^{i\vec{k}_f \cdot \vec{R}} \phi_{1s}(r_{1A})\phi_{1s}(r_{2B})$$

Here $\phi_{1s}(r_{1A})$ and $\phi_{1s}(r_{2B})$ are ground state wave functions of the two atoms and $P_{12}$ is the exchange (antisymmetry) operator.

To define static exchange potential, we consider all the four Coulomb interaction terms: the nucleus-nucleus interaction, electron-electron interaction, interaction between nucleus A and electron 2, and the interaction between nucleus B and electron 1 (see Figure 1). We use atomic masses. The discussion about the importance of atomic and nuclear masses are available in the literature [14].

In the present model the static exchange potential is modified including the effect of long-range dipole-dipole interaction. It is defined as

$$\int d\vec{R} \int d\vec{r}_1 \int d\vec{r}_2 [\psi_f^*(\vec{R},\vec{r}_1,\vec{r}_2)\{V_{van}(R)\}\psi_i(\vec{R},\vec{r}_1,\vec{r}_2)]$$

$$= \int d\vec{r}_1 \int d\vec{r}_2 \int d\hat{R} \int_{R=R_0}^{\infty} R^2 dR \left[ \psi_f^*(\vec{R},\vec{r}_1,\vec{r}_2)\{-\frac{C_W}{R^6}\}\psi_i(\vec{R},\vec{r}_1,\vec{r}_2) \right]$$

It should be noted in Figure 2 that when two atoms are far apart i.e. $R \to \infty$, the potential is almost zero, indicating no interaction between the two atoms. When they proceed to each other, the potential starts becoming more and more negative and reaches a minimum value at $R = R_{min}$, i.e. maximum attraction between the two atoms. The attraction gradually decreases as $R \angle R_{min}$, so the potential gradually increases, becomes zero when the interatomic separation is $R_0$. The interatomic potential starts to be sharply positive as $R \angle R_0$ due to strong static Coulomb interaction between the atoms. So the atoms begin repelling each other strongly and can not proceed further towards themselves. As a result, the minimum value of the interatomic separation should be $R_0$. How the value of $R_0$ should be determined is an important question to evaluate the potential. There is no good literature. In hydrogen molecule the internuclear separation is 1.48 a.u. [15], but the molecule formation is possible only when the two electrons are in singlet (antiparallel spins) state. According to definition of Lennard-Jones 6-12 potential $\sigma$ should be equal to $R_0$ and it is 5.3 a.u. [16] for H-H system. However in a cold-atomic system, when the atoms are very slow, the density seems to control the value of $R_0$.

In the present study we select different values of $R_0$ starting from $1.5 a_0$ to $10 a_0$. We calculate the s-wave elastic phase shift and the corresponding cross section at too low energy region starting from $1 \times 10^{-4}$ eV to 0.1 eV for both the singlet and triplet states of the two electrons in the system. One can determine the corresponding scattering lengths using the help of effective range theory discussed below. Here $a_0$ is the Bohr radius. We adapt most accurate values of $C_W$ reported by Mitroy and Bromley [17].

The effective range theory expresses s-wave elastic phase shift as a function of scattering length and projectile energy so that

$$k \cot \delta_0 = -\frac{1}{a} + \frac{1}{2} r_0 k^2$$

where $\delta_0$ is the s-wave elastic phase shift, $\vec{k}$ is the incident momentum, $a$ is the scattering length and $r_0$ is the range of the potential. Accordingly the scattering length is defined as

$$a = -\lim_{k \to 0}[\frac{\tan \delta_0(k)}{k}]$$

The scattering length, $a < 0$ indicates the possibility of no binding in the system. Only the positive scattering length i.e. $a > 0$ indicates the possibility of binding, so Feshbach resonances and BEC. A rapid change in phase-shift by $\pi$ radian, is an indication of the presence of a Feshbach resonance [18,19] i.e. a binding in the system. A very good theoretical efforts are made to study cold-atomic interaction in H-H [20-23] and alkali atom-atom [24] systems, but no effort is made to study the Ps-H system [5]. We apply the present model (MSEM) in Ps-H system.

**Results and discussion:**

Table 1(a). Comparison of static-exchange s-wave elastic singlet phase shift s ( $\delta_0^+$ ) in radian with MSE data varying the interatomic separation ( R ).

| $k^2$ (a.u.) | Static-exchange | MSE with R=10 a.u. | MSE with R=7 a.u. | MSE with R=5 a.u. | MSE with R=4 a.u. | MSE with R=3 a.u. | MSE with R=2.5 a.u. |
|---|---|---|---|---|---|---|---|
| 1 | 2.4570 | 2.4578 | 2.4581 | 2.4594 | 2.4670 | 2.5207 | 2.6038 |
| 2 | 1.9272 | 1.9283 | 1.9293 | 1.9301 | 1.9381 | 2.0054 | 2.1178 |
| 3 | 1.5390 | 1.5398 | 1.5421 | 1.5429 | 1.5470 | 1.6044 | 1.7172 |
| 4 | 1.2393 | 1.2397 | 1.2426 | 1.2455 | 1.2467 | 1.2862 | 1.3898 |
| 5 | 0.9976 | 0.9978 | 1.0000 | 1.0060 | 1.0068 | 1.0313 | 1.1158 |
| 6 | 0.7979 | 0.7982 | 0.7993 | 0.8078 | 0.8111 | 0.8228 | 0.8901 |
| 7 | 0.6311 | 0.6314 | 0.6319 | 0.6407 | 0.6487 | 0.6535 | 0.7004 |
| 8 | 0.4910 | 0.4910 | 0.4916 | 0.4987 | 0.5115 | 0.5163 | 0.5447 |

Table 1(b). Comparison of static-exchange s-wave elastic triplet phase shift s ( $\delta_0^-$ ) in radian with MSE data varying the interatomic separation ( R ).

| $k^2$ (a.u.) | Static-exchange | MSE with R=10 a.u. | MSE with R=7 a.u. | MSE with R=5 a.u. | MSE with R=4 a.u. | MSE with R=3 a.u. | MSE with R=2.5 a.u. |
|---|---|---|---|---|---|---|---|
| 1 | -0.2472 | -0.2454 | -0.2415 | -0.2337 | -0.2254 | -0.2099 | -0.1944 |
| 2 | -0.4891 | -0.4880 | -0.4825 | -0.4688 | -0.4529 | -0.4215 | -0.3903 |
| 3 | -0.7213 | -0.7210 | -0.7167 | -0.6998 | -0.6773 | -0.6303 | -0.5801 |
| 4 | -0.9400 | -0.9398 | -0.9378 | -0.9210 | -0.8936 | -0.8307 | -0.7606 |
| 5 | -1.1432 | -1.1429 | -1.1423 | -1.1287 | -1.0984 | -1.0199 | -0.9286 |
| 6 | -1.3297 | -1.3294 | -1.3289 | -1.3198 | -1.2895 | -1.1968 | -1.0811 |
| 7 | -1.4994 | -1.4993 | -1.4983 | -1.4937 | -1.4661 | -1.3618 | -1.2225 |
| 8 | -1.6526 | -1.6525 | -1.6513 | -1.6492 | -1.6271 | -1.5157 | -1.3514 |

We reproduce the earlier data [7] using the static-exchange approximation for the values of $k^2$ equal to 1, 2, 3, 4, 5, 6, 7, 8. We calculate the same quantities using the present MSEM code and varying the interatomic separation. A comparison is made of the MSEM data with the reproduced static-exchange data for the s-wave elastic singlet (+) phase shifts in Table 1(a) and triplet phase shifts (-) in Table 1(b). The data are showing a very good agreement with the existing physics [25], all the phase shift results are gradually increasing with the increase of strength of interaction as R decreases.

In Figure 3, the s-wave elastic phase shifts for both the singlet and triplet channels using present theory (MSEM) are presented and compared with the static exchange model data at too low energy region. The corresponding s-wave elastic cross sections are presented in Figure 4a for the singlet channel and in Figure 4b for the triplet channel. The different minimum values of interatomic separation are chosen as $R_0 = 3a_0, 5a_0, 7a_0$ and $10a_0$ respectively to define the modified long-range van der Waals potential.

To evaluate the scattering lengths the effective range theory is useful. $k \cot \delta_0$ is plotted against $k^2$, to evaluate $-1/a$ in Figure 5a for singlet channel and in Figure 5b for triplet channel. The scattering length vary systematically with varying minimum values of interatomic separation chosen as $3a_0, 5a_0, 7a_0$ and $10a_0$. When $R_0 = 10a_0$ or greater, all the data almost coincide with the

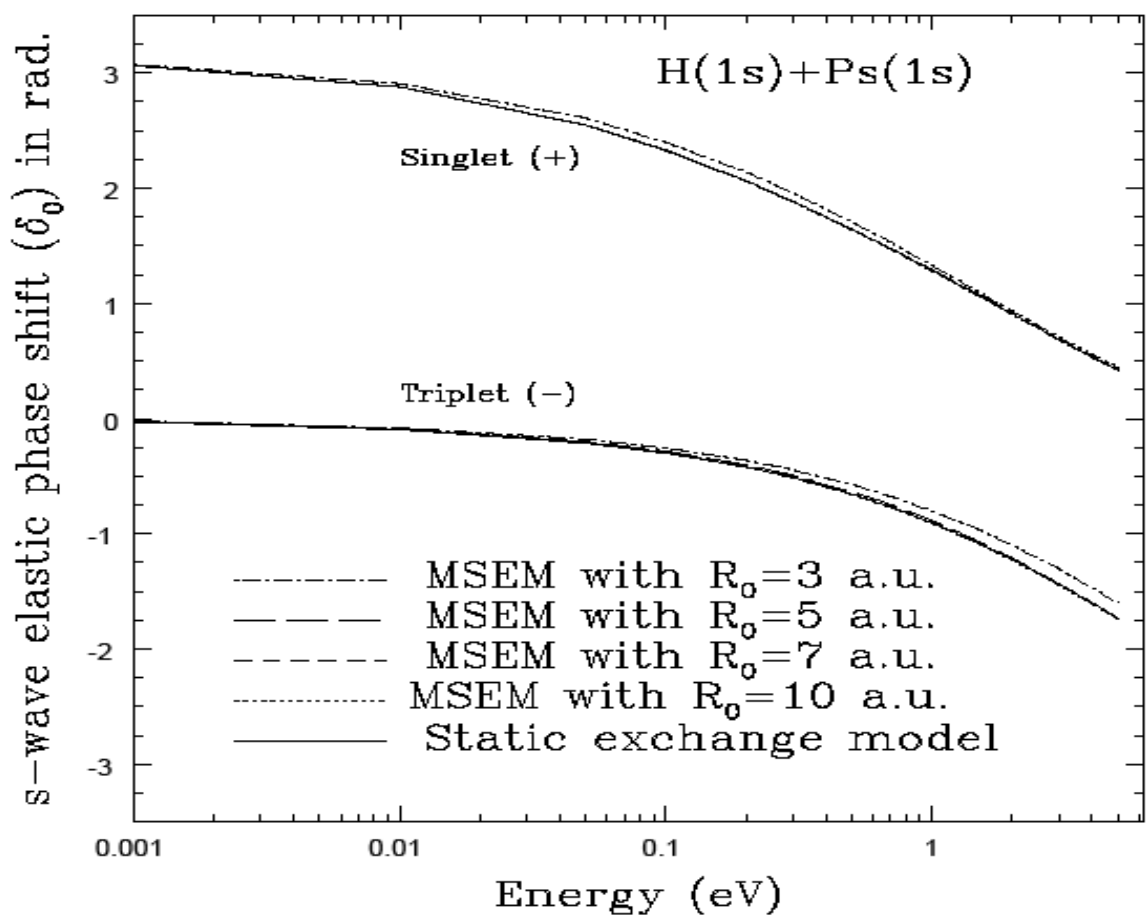

Figure 3

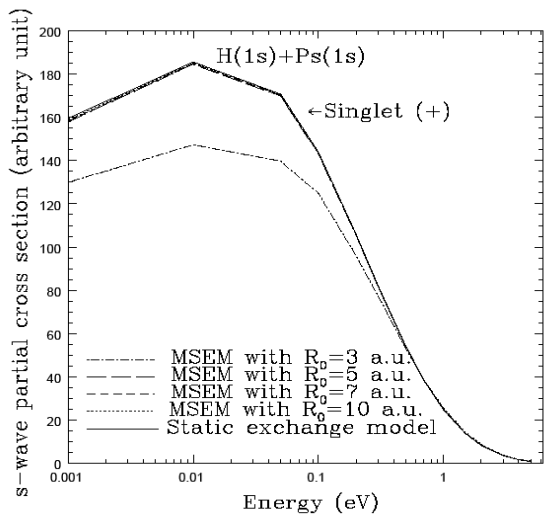

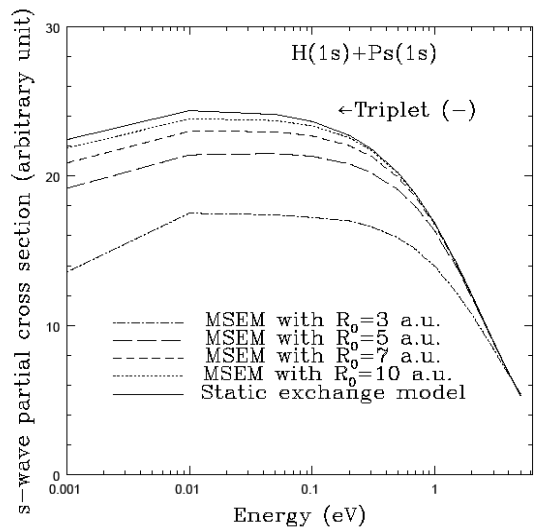

Figure 4a

Figure 4b

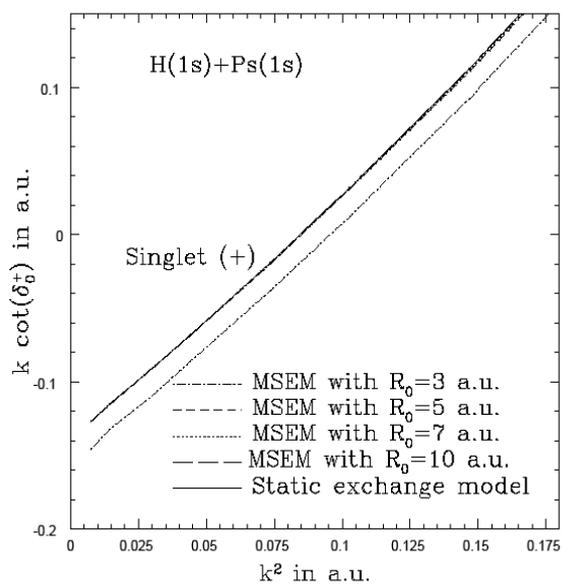

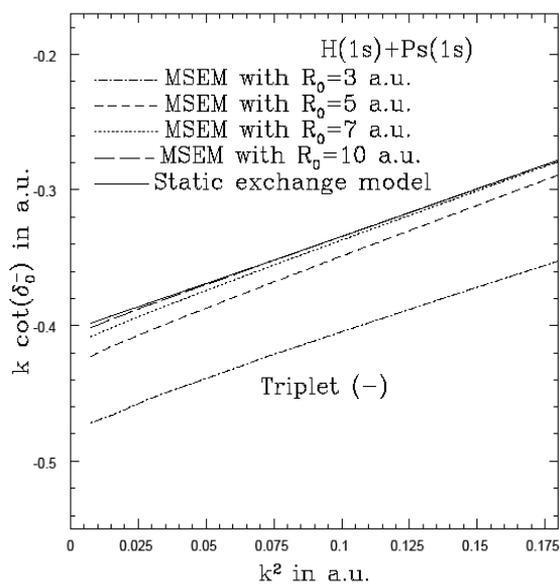

Figure 5a

Figure 5b

SEM data. There are always a very small difference between the SEM data and MSEM data almost upto $R_0 = 50a_0$ indicating extent of long-range potential. Here all the scattering lengths are positive. It indicates the possibility of binding and Feshbach resonances. The variation of triplet scattering length is more sensitive to long-range interaction than singlet scattering length. A very large number of mesh points are required to study the resonances. The same set of input data points are used to draw all the curves discussed; the tabular data using SEM for s-wave elastic phase shifts and cross sections are presented in Table 2.

Table 2. The s-wave elastic phase shifts and cross sections for Ps-H system using SEM theory

| Energy (eV) | Partial wave (l) | Singlet phase shift (rad) | Triplet phase shift (rad) | Singlet cross section (arbitrary unit) | Triplet cross section (arbitrary unit) |
|---|---|---|---|---|---|
| 1.0E-4 | 0 | 0.308E+01 | -0.926E-02 | 0.101E+04 | 0.233E+02 |
| 1.0E-3 | 0 | 0.306E+01 | -0.287E-01 | 0.159E+03 | 0.224E+02 |
| 1.0E-2 | 0 | 0.288E+01 | -0.948E-01 | 0.186E+03 | 0.244E+02 |
| 1.0E-1 | 0 | 0.233E+01 | -0.299E+00 | 0.144E+03 | 0.236E+02 |
| 1.0E+0 | 0 | 0.128E+01 | -0.905E+00 | 0.250E+02 | 0.168E+02 |
| 2.0E+0 | 0 | 0.908E+00 | -0.122E+01 | 0.846E+01 | 0.120E+02 |
| 3.0E+0 | 0 | 0.687E+00 | -0.144E+01 | 0.365E+01 | 0.891E+01 |
| 4.0E+0 | 0 | 0.535E+00 | -0.160E+01 | 0.176E+01 | 0.679E+01 |
| 5.0E+0 | 0 | 0.421E+00 | -0.173E+01 | 0.908E+00 | 0.530E+01 |

In addition, we study the Feshbach resonances at very low energy region $10^{-4}$ to $10^{-2}$ for triplet channels varying the values of $R_0$ and using a very fine mesh-points. Very interesting features and Feshbach resonances are observed. The s-wave elastic phase shifts are displayed in figure 6(a), 6(b), 6(c), 6(d), 6(e), 6(f), 6(g), 6(h) using static exchange model (SEM) and present MSEM with $R_0$ =6, 5, 4, 3, 2.5, 2.3, 2.2 a.u. respectively. The corresponding cross sections appear as Figures 7(a), 7(b), 7(c), 7(d), 7(e), 7(f), 7(g) and 7(h). A very narrow Feshbach resonance at E = $1.25 \times 10^{-4}$ eV continues its presence when $R_0$ varies from 5 to 2.5 a.u. There is a dip in cross section when $R_0$ =5 a.u. but that dip transforms into a peak when $R_0$ is less than 5 a.u. Again when $R_0$ =2.5 a.u. the cross section curve shows almost similar behaviour with SEM data. The observed interesting behaviour of phase shifts and cross sections motivate the theoretical findings if the two atoms approach more close to each other. Accordingly the calculations are made using $R_0$ = 2.1, 2.0, 1.75 and 1.5 a.u. and are presented the phase shifts in Figure 8(a), 8(b), 8(c), 8(d) and the cross sections in 9(a), 9(b), 9(c), 9(d). The earlier resonance at $1.25 \times 10^{-4}$ eV disappears and more new resonances appear.

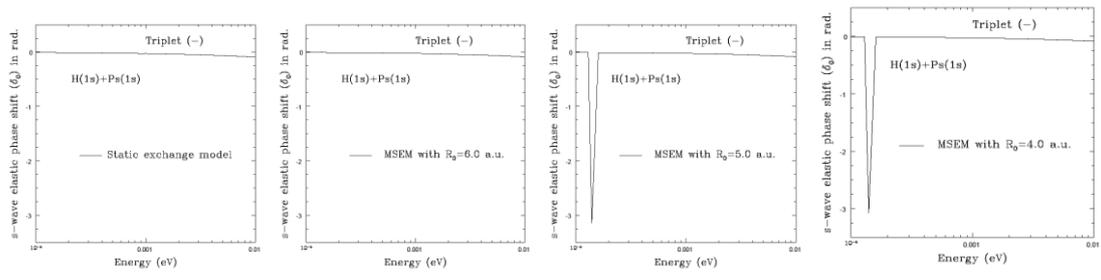

Figure 6(a)  Figure 6(b)  Figure 6(c)  Figure 6(d)

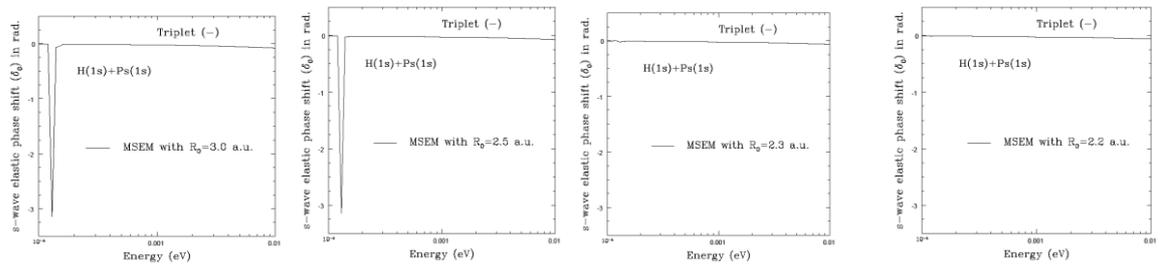

Figure 6(e)  Figure 6(f)  Figure 6(g)  Figure 6(h)

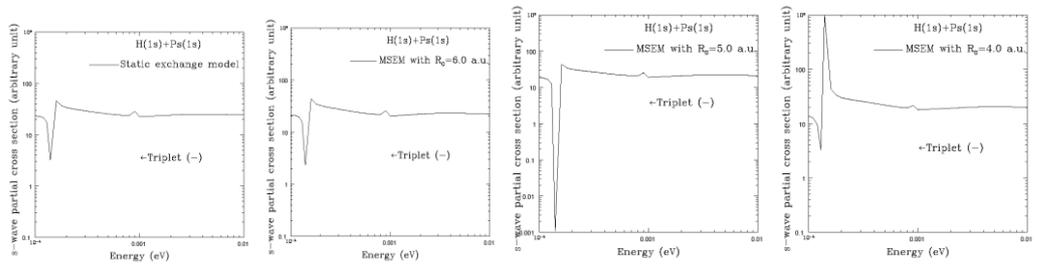

Figure 7(a)  Figure 7(b)  Figure 7(c)  Figure 7(d)

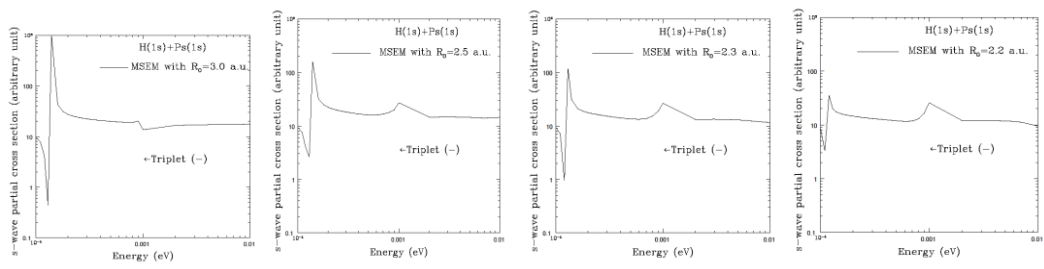

Figure 7(e)  Figure 7(f)  Figure 7(g)  Figure 7(h)

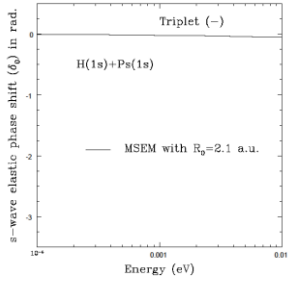 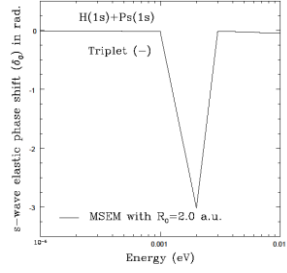 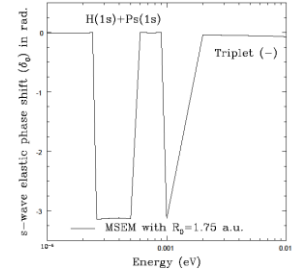 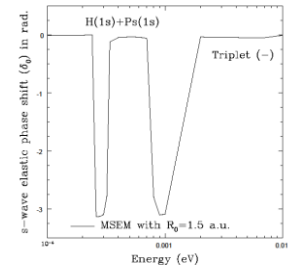

        Figure 8(a)            Figure 8(b)           Figure 8(c)           Figure 8(d)

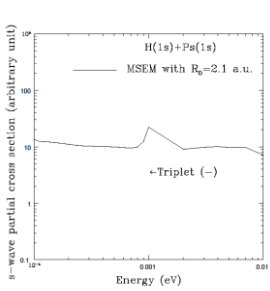 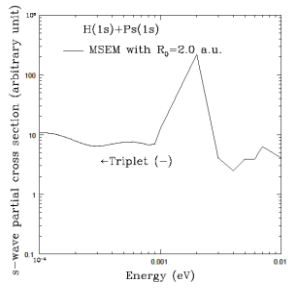 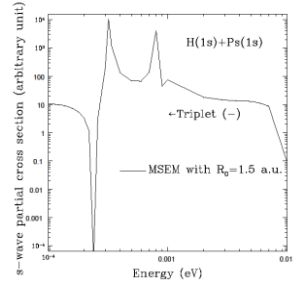 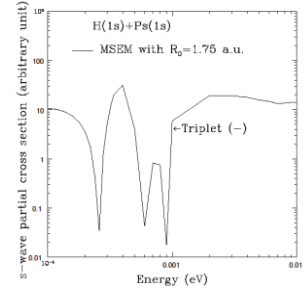

        Figure 9(a)            Figure 9(b)           Figure 9(c)           Figure 9(d)

**Conclusion:**

A modified static exchange model (MSEM) is introduced to study low energy atom and atom scattering. It could be useful to control and/or tune the elastic s-wave scattering length in ultracold atomic collisions. We apply the model in positronium (Ps) and hydrogen (H) system when both the atoms are in ground states. Since both are hydrogen like and one of them Ps has very light mass and strong polarizability, the present system is very useful to extract the basic physics very accurately. We study the s-wave elastic phase shift and the corresponding cross section in the energy range $10^{-4}$ eV to 0.1 eV using the present model (MSEM) and the static exchange model (SEM). The parameter $R_0$ appears to be the determining factor to occur Feshbach resonances in the system. As the atoms are very slow, the magnitude of $R_0$ should be controlled by the density of the atoms in optical lattices.

**Acknowledgement:**

Author is thankful to Department of Science & Technology (DST) India for the full financial support through grant no. SR/WOSA/PS-13/2009. She is grateful to Gleb Gribakin, Allen P. Mills and R. J. Drachman for reading the manuscript and their valuable comments.